\documentclass{appolb}
\RequirePackage[T1]{fontenc}
\usepackage{graphicx}
\usepackage{amssymb}
\usepackage{amsfonts}

\begin{document}
\title{ELECTRICITY REAL OPTIONS VALUATION}
\author{Ewa Broszkiewicz-Suwaj
\address{Hugo Steinhaus Center, Institute of Mathematics and Computer Science \\Wroc{\l}aw University of Technology\\ Wyspia\'nskiego 27, 50-370 Wroc{\l}aw, Poland\\ \ttfamily{Ewa.Broszkiewicz-Suwaj@pwr.wroc.pl}}} 
\maketitle
\begin{abstract}
In this paper a real option approach for the valuation of real assets is presented. Two continuous time models used for valuation are described: geometric Brownian motion model and interest rate model. The valuation for electricity spread option under Vasicek interest model is placed and the formulas for parameter estimators are calculated. The theoretical part is confronted with real data from electricity market.
\end{abstract}
\section{Introduction}

The liberalization of electricity market caused that modeling on this market became very important skill. It helps us to minimize loss and hedge our position. It is a very interesting fact that spread option  could be used for valuation some real assets as power plants or transmission lines. But before that we need to know the spread option price formula. Very popular model used for option valuation is the geometric Brownian motion model but it is not very efficient. In the 2005 the idea of modeling domestic electricity market using interest rate model was introduced by Hinz, Grafenstein, Verschue and Wilhelm \cite{hinz}. They valuated European call option written on power forward contract under Heath Jarrow Morton model and, in this way, created very interesting class of models.

The aim of my work is the valuation and calibration of electricity spread option under interest rate model applied for electricity market. I start with assumption of Vasicek model and using martingale methodology \cite{musiela} valuate spread option. 
Using the maximum likelihood function methodology I estimate model parameters. I compare constructed model with geometric Brownian motion model by applying both models to real option valuation. I make simulations to show the difference between to discussed models.

My paper is organized in the following way. At the beginning (Section 2) I describe what does it mean that we bought a spread option, next (Section 3) I introduce the reader into real options world. In Section 4 I describe valuation  methodology for spread option under  interest rate model and present also option price formula  for geometric Brownian motion model. The calibration methods for both models are described in Section 5. At the end, in Section 6, all theoretical deliberation are confronted with real data and some simulation results are presented.

\section{Electricity Spread Options}
In this section two interesting cross commodity derivatives on electricity market are described. 
The first one is the spark spread  option, which is based on fact that some power plants convert gas into electricity. The underlying instrument is the difference between the gas and electricity prices (the spark spread). The basic parameter connected with this kind of instrument is the heat rate, the ratio which describes the amount of gas required to generate 1 MWh of electricity.
 The definition of such an instrument has a form \cite{djs, wer}: 
 An European spark spread call option  written on fuel $F$, at fixed swap ratio $K$, gives its holder the right,  but not the obligation to pay K times the unit price of fuel $F$  at the options maturity $T$ and receive the price of one unit of electricity.

It is easy to imagine such kind of option which better fits the Polish electricity market. We should assume that the underlying instrument is the difference between the carbon and electricity prices. But in this time there is no possibility of valuation of such an option because we don't have the representative carbon price.
Generally, if we assume that $P_E$ and $P_F$ are respectively future price of 1MWh of electricity and the future price of the unit of fuel and K is the swap ratio than we could describe the payoff of the European electricity-fuel spread call option as
\[
C_F(P_E,P_F,T)=\max[P_E(T)-KP_F(T),0]
\]
and  the payoff of the European electricity-fuel spread put option has form
\[
P_F(P_E,P_F,T)=\max[KP_F(T)-P_E(T),0]
\]

The second derivative is the locational spread option. It is based on fact that transmission of power from one location to another is very popular transaction. It is normal, for transmission system, that the power is moved from the place of lower price  to the place of higher price and this is why the transaction is profitable. The whole transaction depends on the difference between the electricity prices and also on delivery costs and for hedging we could use options. This kind of instrument could be defined in following way \cite{djs}:
An European call option on the locational spread between the location one and location two, with maturity $T$, gives its holder the right  but not the obligation to pay the price of one unit of electricity at location one at time T and receive the price of $K$ units of electricity at location two. 
Assume that $P_1$ and $P_2$ are the electricity prices at the first location and second location respectively. The payoff of the European locational spread call option is given by
\[
C_L(P_1,P_2,T)=\max[P_1(T)-KP_2(T),0]
\]
The put option is defined similar and  the payoff of the European electricity-fuel spread put option has form
\[
P_L(P_1,P_2,T)=\max[KP_2(T)-P_1(T),0]
\]

\section{Real Options}
Suppose that for describing the commodity we use three qualities  $(G,t,L)$: G - the nature of good, t - time when it is available, L - location where it is available.
We could define \cite{ron} a real option as  technology to physically convert one or more input commodities $(G,t,L)$ into an output commodity $(G',t',L')$. For example most of power plants are real option because they give us the right to convert fuel  into electricity. The transmission line is also real option. It gives us the right to change the electricity in one location into electricity in second location. The works of Deng, Johnson and Sogomonian \cite{deng, djs} contain two formulas defining how to valuate generation and transmission assets.
If we define that $u_F(t)$ is a one unit of the time-t right to use generation asset we could say that it is the value of just maturing, time-t call option on the spread between electricity and fuel prices $C_F(t)$ and the one unit  value of capacity of power plant using some fuel $F$ is given by 
\[
V_F=\int_0^Tu_F(t)dt=\int_0^TC_F(t)dt
\]
where $T$ is the length of power plant life.

Similar if we define that $u_L^{AB}(t)$ is a one unit  the time-t right to convert one unit of electricity in location A into one unit of electricity in location B  we could say that it is the value  of just maturing,time-t call option on the spread between electricity prices in location A and B $C_L^{AB}(t)$. The one unit  value of such transmission asset is given by 
\[
V_L=\int_0^Tu_L^{AB}(t)dt+\int_0^Tu_L^{BA}(t)dt=\int_0^TC_L^{AB}(t)dt+\int_0^TC_L^{BA}(t)dt
\]
where $T$ is the length of transmission network  life.

\section{Valuation methods}
In this section I present the widely known geometric Brownian   motion model and I valuate the call spread option for the new, interest rate model using martingale methodology. All calculations are described below.

\subsection{Geometric Brownian Motion Model }

Suppose that the future prices of commodity are described by following stochastic differential equations
\[
dP_1(t,T)=\mu _1P_1(t,T)dt+\sigma _1P_1(t,T)dW_{t,1} ,
\] 
\[
dP_2(t,T)=\mu _2P_2(t,T)dt+\sigma _2P_2(t,T)dW_{t,2} ,
\]
where $W_{t,1}=\rho W_{t,2}+\sqrt{1-\rho^2}W_{t,2}^{'}$ and $W_{t,2}$, $W_{t,2}^{'}$ are i.i.d. Brownian motions. It is known fact \cite{ron},  that the price of the spread call option with swap ratio $K$ and time to maturity $T$, written on futures contract with maturity $U<T$  is given by
\[
C_{1}(t)=e^{-r(T-t)}[P_1(t,U)\Phi(d_+(t))-KP_2(t,U)\Phi(d_-(t))],
\]
where
\[
d_\pm(t)=\frac{ln\frac{P_1(t,U)}{KP_2(t,U)}\pm\frac{\sigma^2(T-t)}{2}}{\sigma\sqrt{T-t} }.
\]
and
\[
\sigma^2=\sigma_1^2-2\sigma_1\rho\sigma_2+\sigma_2^2.
\]

\subsection{Interest Rate Model}
For domestic currency, for example MWh, we denote two processes: $p_{1}(t,T)$,
$p_{2}(t,T)$ which are the future prices of one unit of commodity. The interest rate functions for such processes are respectively
\[
dr_{t,1}=(a_1-b_1r_{t,1})dt+\sigma_1dW_{t,1}
\]
and
\[
dr_{t,2}=(a_2-b_2r_{t,2})dt+\sigma_2dW_{t,2},
\]
where $W_{t,1}=\rho W_{t,2}+\sqrt{1-\rho^2}W_{t,2}^{'}$ and $W_{t,2}$, $W_{t,2}^{'}$ are i.i.d. Brownian motions.
We assume that there exist the savings security $N_t$ (for example in USD), with constant interest rate $r$, for which 
\[
P(t,T)=\frac{p(t,T)}{e^{-rt}N_t}
\]
is the USD price of future delivery of 1 unit  of commodity.
We know \cite{hinz} that  there exist a martingale measure $\mathbb{P}$ for which the discounted processes $\frac{p_{1}(t,T)}{N_{t}}$,
$\frac{p_{2}(t,T)}{N_{t}}$ are martingales.
We have
\[
C_{2}(0)=N_{0}E_\mathbb{P}((p_1(T,U)-Kp_2(T,U))^+N_T^{-1}|F_{0})/e^{-rt}N_{t}
\]
We  define the new  discounting processes for i=1,2 as
\[
dB_{t,i}=B_{t,i}r_{t,i}dt,
\]
where $B_{0,i}=1.$
If we change the measure from $\mathbb{P}$ to $\mathbb{P}_1$
\[\frac{d\mathbb{P}}{d\mathbb{P}_1}=\frac{N_T B_{0,1}}{N_0 B_{T,1}}
\]
we know that processes $\tilde{p_1}(t,T)=\frac{p_{1}(t,T)}{B_{t,1}}$, $\tilde{p_2}(t,T)=\frac{p_{2}(t,T)}{B_{t,1}}$  are  $\mathbb{P}_1$-martingales.  
From interest rate theory we obtain
\[
d\tilde{p_1}(t,T)=\tilde{p_1}(t,T)n_1(t,T)dW_{t,1},
\]
where
\begin{equation}
n_1(t,T)=-\frac{\sigma_1}{b_1}(1-e^{-b_1(T-t)}).
\label{n1}
\end{equation}
If we change the measure again from $\mathbb{P}_1$ to $\mathbb{P}_2$ 
\[
\frac{d\mathbb{P}_1}{d\mathbb{P}_2}=\frac{B_{T,1}B_{0,2}}{B_{0,1}B_{T,2}}
\]
we know that
$\hat{p_2}(t,T)=\frac{p_{2}(t,T)}{B_{t,2}}$ and
$\hat{p_1}(t,T)=\frac{p_{1}(t,T)}{B_{t,2}}=\frac{\tilde{p_1}(t,T)B_{t,1}}{B_{t,2}}$  are  $\mathbb{P}_2$-martingales and
similar to the earlier situation, we have 
\begin{equation}
d\hat{p_2}(t,T)=\hat{p_2}(t,T)n_2(t,T)dW_{t,2},
\label{p2}
\end{equation}
where
\begin{equation}
n_2(t,T)=-\frac{\sigma_2}{b_2}(1-e^{-b_2(T-t)}).
\label{n2}
\end{equation}
After simple calculations we also have
\begin{equation}
d\hat{p_1}(t,T)=\hat{p_1}(t,T)n_1(t,T)d\tilde{W}_{t,1},
\label{p1}
\end{equation}
where $\tilde{W}_{t,1}=\rho W_{t,2}+\sqrt{1-\rho^2}\tilde{W}_{t,2}^{'}$, and $\tilde{W}_{t,2}^{'}={W}_{t,2}^{'}+\frac{r_{t,1}-r_{t,2}}{n_1(t,T)\sqrt{1-\rho^2}}t$.
We also assume that
\begin{equation}
d\hat N_t=\hat N_t v dV_t,
\label{n}
\end{equation}
where $ V_t=\rho_1W_{t,2}+\sqrt{1-\rho_1^2}W_{t,2}^{''},$ and $W_{t,2}$, $W_{t,2}^{'}$ ,  $W_{t,2}^{''}$ are independent Wiener processes.
For discounted processes following equation is true

\begin{equation}
P_i(t,T)=\frac{p_i(t,T)}{e^{-rt}N_t}=\frac{\hat p_i(t,T)}{e^{-rt}\hat N_t}.
\label{zal}
\end{equation}

Having the necessary stochastic differential equations we could price the option.
We change the measure from $\mathbb{P}_2$ to $\mathbb{Q}$  in following way 
\[
\frac{d\mathbb{P}_2}{d\mathbb{Q}}=\frac{B_{T,2}p_2(0,U)}{B_{0,2}p_2(T,U)}.
\]
Process $X(t,T)=\frac{p_1 (t,T)}{p_2 (t,T)}=\frac{\hat{p_1}(t,T)}{\hat{p_2}(t,T)}$ is $\mathbb{Q}$-martingale. 
From Ito lemma we know that 
\[
dX(t,T)=X(t,T)(n_1(t,T)d\hat{W}_{t,1}-n_2(t,T)dW_{t,2}),
\]
where $\hat{W}_{t,1}=\rho W_{t,2}+\sqrt{1-\rho^2}\hat{W}_{t,2}$ and $\hat{W}_{t,2}=\tilde{W}_{t,2}^{'}+\frac{n_2^2(t,T)-n_2(t,T)\rho n_1(t,T)}{n_1(t,T)\sqrt{1-\rho^2}}t$.
Now, for calculation of the option price,  we could use the Black-Scholes formula 
\[
C_{2}(0)=\frac{p_2(0,U)e^{r0}}{N_0}E_\mathbb{Q}((\frac{p_1(T,U)}{p_2(T,U)}-K)^+|F_{0})=P_2(0,U)E_\mathbb{Q}((\frac{P_2(T,U)}{P_1(T,U)}-K)^+|F_{0})= 
\]
\[
P_2(0,U)E_\mathbb{Q}((X(T,U)-K)^+|F_{0})=P_2(0,U)(X(0,U)\Phi(d_+)-K\Phi(d_-))=
\]
\[
P_1(0,U)\Phi(d_+)-KP_2(0,U)\Phi(d_-),
\]
where
\[
d_\pm=\frac{ln\frac{X(0,U)}{K}\pm\frac{\sigma^2(0,U)}{2}}{\sigma(0,U) },
\]
\[
\sigma^2(t,T) =\int_t^T(n_1^2(u,T)-2n_2(u,T)\rho n_1(u,T)+n_2^2(u,T))du,
\]
and $\Phi$ is the normal  cumulative distribution function.
For every time point $0\leq t \leq T$ the option price with swap ratio $K$ and time to maturity $T$, written on futures contract with maturity $U<T$ is given by
\[
C_{2}(t)=P_1(t,U)\Phi(d_+(t))-KP_2(t,U)\Phi(d_-(t)),
\]
where
\[
d_\pm(t)=\frac{ln\frac{P_1(t,U)}{KP_2(t,U)}\pm\frac{\sigma^2(t,U)}{2}}{\sigma(t,U) }.
\]
This methodology could be used directly for locational spread options  and also for fuel-electricity spread options if we assume that the swap ratio between MWh and unit of fuel is one.

\section{Historical Calibration }
In this section we describe how to fit our models for real, historical data. 
At the beginning we assume that we are given historical prices of future contracts $P_1(t_k,T_j)$ and $P_2(t_k,T_j)$,  $k=0,\ldots,n$, $j=0,\ldots,m$, in discrete time points $t_0<t_1<\ldots<t_n$ and $T_0<T_1<\ldots<T_m$, where $t_{k+1}-t_k=dt$ and $T_{j+1}-T_j=\Delta T$ . 

For geometric Brownian motion model the calibration methodology is not very complicated. We analyze the returns of future prices of instrument and $\mu$ is its mean, $\sigma^2 $ is its variance and correlation parameter is simply the correlation between returns of two instruments. But for interest rate model the calibration is quite complicated, especially for multidimensional HJM model  \cite{b}. Calibration for discussed Vasicek model is presented below.
\\
Let us consider following process 
\[
\eta_i(t, T_j)=\frac{\hat p_i(t,T_j)}{\hat p_i(t,T_{j+1})}=\frac{P_i(t,T_j)}{P_i(t,T_{j+1})}.
\]
From It\^o lemma we know that

\begin{flushleft}
$d\eta_i(t, T_j)=\eta_i(t, T_j)[(n_i^2(t,T_{j+1})-n_i(t,T_{j+1})n_i(t,T_{j}))dt+$
\end{flushleft}
\begin{flushright}
$+(n_i(t,T_{j})-n_i(t,T_{j+1}))dW_{t,i}]$
\end{flushright}
We could write that $n_i(t,T)=n_i(T-t)$ because this function depends only from the difference between the maturity time $T$ and the time point $t$. If we then consider the process
\[
s_i(T-t)=\frac{d\eta_i(t, T)}{\eta_i(t, T)},
\]
we know that  $s_i(T-t)$ is normally distributed  with mean 
\[ \alpha_i(T-t)=(n_i^2(T+\Delta T-t)-n_i(T+\Delta T-t)n_i(T-t))dt \]
 and variance
\[ \beta^2_i(T-t)=(n_i(T-t)-n_i(T+\Delta T-t))^2dt .\]
Knowing the form of functions $n_i(T-t)$ (\ref{n1}),(\ref{n2}) we see that
\begin{equation}
 \beta^2_i(T-t)=\left(\frac{\sigma_i}{b_i}e^{-b_i(T-t)}[1-e^{-b_i\Delta T}]\right)^2dt
 \label{bb}
\end{equation}
and
\begin{equation}
\frac{\beta^2_i(T-t)}{\beta_i^2(T-t+dt)}=e^{2b_idt}
\label{b}
\end{equation}
After discretisation and for assumption that $dt=\Delta T=1$, $T-t=p\Delta T$ and $j=1,\ldots, m$ we could say that the estimator of $\beta$ has the form 
\[
\hat{\beta_i}^2(p\Delta T)=\frac{1}{m}\sum_{j=1}^m(s_{i,j}^2(p\Delta T)-s_{i,j}(p\Delta T)\bar{s_i}),
\]
where
\[
\bar{s_i}=\frac{1}{m}\sum_{j=1}^ms_{i,j}(p\Delta T)
\] 
and  we put
\[
s_{i,j}(p\Delta T)=\frac{\frac{P_i(T_j-p\Delta T,T_j)}{P_i( T_j-p\Delta T,T_j+\Delta T)}-\frac{P_i(T_j-(p+1)\Delta T,T_j)}{P_i(T_j-(p+1)\Delta T,T_j+\Delta T)}}{\frac{P_i(T_j-p\Delta T,T_j)}{P_i(T_j-p\Delta T,T_j+\Delta T)}}.
\]
So using (\ref{bb}),(\ref{b}) we have
\[
\hat b_i=\frac{1}{2\Delta T}\ln{\frac{\hat{\beta_i}^2(p\Delta T)}{\hat{\beta_i}^2((p+1)\Delta T)}},
\]
\[
\hat \sigma_i=\frac{\hat{\beta_i}^2(p\Delta T)\hat b_i} {e^{-\hat b_i(p\Delta T)}[1-e^{-\hat b_i\Delta T}]}.
\]
It is easy to notice that the correlation parameter between processes $s_1(T-t)$ and $s_2(T-t)$ is $\rho$, so we have 
\[
\hat \rho=\frac {\sum_{j=1}^m(s_{1,j}(p\Delta T)-\bar{s_1})(s_{2,j}(p\Delta T)-\bar{s_2})}   {\sqrt{\sum_{j=1}^m(s_{1,j}(p\Delta T)-\bar{s_1})^2}\sqrt{\sum_{j=1}^m(s_{2,j}(p\Delta T)-\bar{s_2})^2}}.
\]
At the end we should calculate also parameters connected with process $N_t$. 
From equation (\ref{zal}) we know that for i=1,2
\[
\xi_{i}(t,T)=\frac{\hat p_i(t,T)}{\hat N_t}=e^{-rt}P_i(t,T).
\]

Using It\^o lemma and formulas  (\ref{n}), (\ref{p1}), (\ref {p2}) we could calculate following dynamic

\begin{flushleft}
$
d\xi_{2}(t,T)=\xi_{2}(t,T)[v^2-vn_2(t,T)\rho_1]dt+\xi_{2}(t,T)[(n_2(t,T)-v\rho_1)dW_{t,2}-
$
\end{flushleft}
\begin{flushright}
$
-v\sqrt{1-\rho_1^2}dW_{t,2}''].
$
\end{flushright}
We know that the process
\[
y(T-t)=\frac{d\xi_{2}(t,T)}{\xi_{2}(t,T)}
\]
is normally distributed with mean $(v^2-vn_2(t,T)\rho_1)dt$ and variance $(n_2^2(t,T)-2vn_2(t,T)\rho_1+v^2)dt$ so we have that for $dt=1$
\[
\hat v^2=\hat n_2^2(p\Delta T)+2\bar{y}-\frac{\sum_{j=1}^{m}(y_j(p\Delta T)-\bar{y})^2}{m-\sum_{j=1}^{m}(y_j(p\Delta T)-\bar{y})}
\]
and
\[
\hat \rho_1=\frac{\hat v^2-\bar{y}}{\hat v\hat n_2(p\Delta T)}
\]
where
\[
y_j(p\Delta T)=\frac{e^{-r\Delta T}P_i(T_j-p\Delta T,T_j)-P_i(T_j-(p+1)\Delta T,T_j)}{P_i(T_j-(p+1)\Delta T,T_j)\Delta T}
\]
and
\[
\bar y=\frac{1}{m}\sum_{j=1}^{m}y_j(p\Delta T).
\]

\section{Simulation and Conclusion}
For simulation I used  data from New York Mercantile Exchange (NYMEX). I considered historical quotation of future natural gas (Henry Hub) and electricity (PJM) contracts since January, 2004  until March, 2006. The parameters were calculated using calibration methods described before. All estimated parameters are presented in Table \ref{t1}. I assumed that the constant interest rate is $r= 0.05$. For valuation of gas fired power plant I assumed that the life-time of the power plant is $T=15$ years and $P_{E,0}=55.750$ USD, $P_{F,0}=6.3080$ USD.  
\begin{table}[h]
\begin{flushright}
TABLE I
\end{flushright}
{Estimated parameters for GBM model and for interest rate model using historical data from New York Mercantile Exchange. }
\begin{center}

\begin{tabular}{l|l}
\hline
\hline
Geometric Brownian Motion & Interest Rate Model\\
\hline
$\sigma_e $ \hskip 1cm 1.0945 & $\sigma_e $ \hskip 1cm 0.0678 \\

$\sigma_g$ \hskip 1cm 1.2943 & $\sigma_g $ \hskip 1cm 0.0042\\

$\mu_e$ \hskip 1cm 4.4098& $b_e$ \hskip 1cm  3.7515\\

$\mu_g$ \hskip 1cm 4.8145 &$b_g $ \hskip 1cm 1.8205\\

$\rho$ \hskip 1.2cm 0.8688 & $\rho $ \hskip 1.1cm 0.1892\\

& $\rho_1$\hskip 1cm  0.7266\\
& $v$ \hskip 1.1cm 0.0668\\

\end{tabular}

\label{t1}
\end{center}
\end{table}

\begin{figure}[h]
	\begin{center}
		\includegraphics[width=11cm]{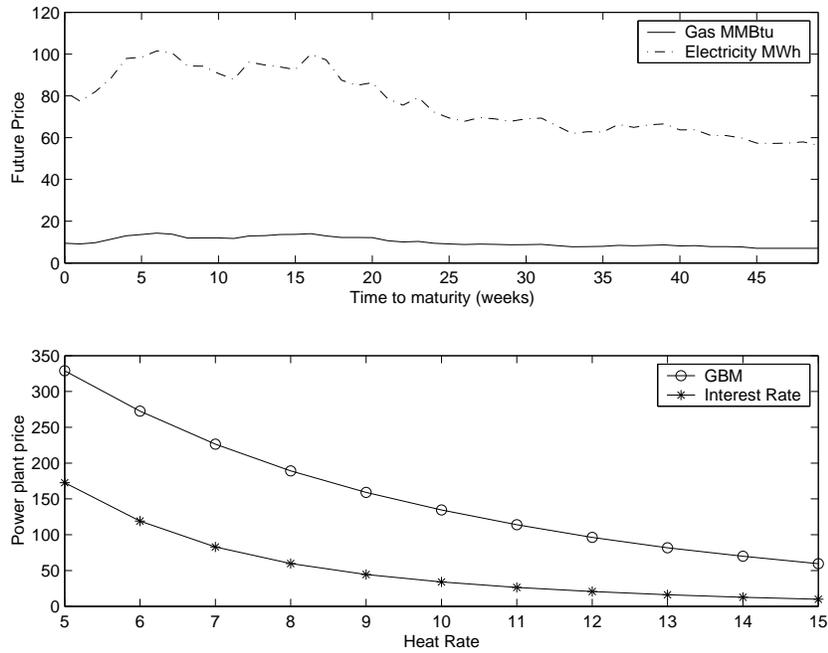}
	\end{center}
	\label{f111}
	\caption{\textit{Top panel:} Future prices of natural gas and electricity for contract maturing in March, 2006. \textit{Bottom panel:} Simulated unit value of gas fired power plant, with life length 15 years, for both models. }
\end{figure}

In Figure 1. we see the value of power plant for the heat rate ranging from 5 to 15 for both presented models. We could notice that  there is difference in changes dynamic for analyzed models. The value of power plant for interest rate model  is much more smaller than for GBM model and it tends to zero when the heat rate goes up. It is a very good feature, because in reality the value of power plant for heat rate greater than $\frac{P_{E,0}}{P_{F,0}}\approx 9$ should be close to zero. Looking at work of Deng we could say, that the value of power plant under GBM model is usually too high, so also in this aspect the interest rate model gives better results.

\end{document}